# Effect of temperature on $In_xGa_{1-x}As/GaAs$ quantum dot lasing

## Mahdi Ahmadi Borji[1] and Esfandiar Rajaei[2]

Department of Physics, The University of Guilan, Namjoo Street, Rasht, Iran

## ABSTRACT

*In this paper, the strain, band-edge, and energy levels of pyramidal $In_xGa_{1-x}As/GaAs$ quantum dot lasers (QDLs) are investigated by 1-band effective mass approach. It is shown that while temperature has no remarkable effect on the strain tensor, the band gap lowers and the radiation wavelength elongates by rising temperature. Also, band-gap and laser energy do not linearly decrease by temperature rise. Our results appear to coincide with former researches.*

## I.     INTRODUCTION

Semiconductor lasers are the most important lasers that are used in cable television signals, telephone and image communications, computer networks and interconnections, CD-ROM drivers, reading barcodes, laser printers, optical integrated circuits, telecommunications, signal processing, and a large number of medical and military applications. Quantum dot semiconductor lasers due to the discrete density of states have low threshold current and temperature dependence, high optical gain and quantum efficiency and high modulation speed are superior to other lasers. Effects of various factors such as temperature (Chen and Xiao, 2007, Kumar et al., 2015, Narayanan and Peter, 2012, Rossetti et al., 2009), size of the quantum dots (Baskoutas and Terzis, 2006, Pryor, 1998), stoichiometric percentage of constituent elements of the laser active region (Shi et al., 2011), substrate index (Povolotskyi et al., 2004), strain effect (Pryor and Pistol, 2005, Shahraki and Esmaili, 2012), wetting layer (WL), and distribution and density of quantum dots are shown to be important in the energy levels and performance of QDLs. Therefore, finding the functionality of the impact of these factors can help in optimization of the performance of quantum dot lasers. Temperature effects should be paid attention, since any adverse effects that happen by change in the laser usage conditions should be forecasted.

1 Email: Mehdi.p83@gmail.com

2 Corresponding author, E-mail: Raf404@guilan.ac.ir

QD nanostructures have been the focus of many investigations due to their optical properties arising from the quantum confinement of electrons and holes (Markéta ZÍKOVÁ, 2012, Ma et al., 2013, Danesh Kaftroudi and Rajaei, 2010, Nedzinskas et al., 2012). By now, QD materials have found very promising applications in optical amplifiers and semiconductor lasers (Bimberg et al., 2000, Gioannini, 2006, Danesh Kaftroudi and Rajaei, 2011, Asryan and Luryi, 2001). They are very important in new laser devices and solar cells. Therefore, having a ubiquitous view of energy states, strain, and other physical features, and their change by varying some factors such as temperature which affects the lasing process of a QD is instructive. Based on this fact, many research groups attempt to develop and optimize QDs to fabricate optoelectronic devices with better performance.

Finding a way to enhance the efficiency of a QD with fixed size can be helpful. Among many materials, InGaAs/GaAs devices are faced by many scientists due to their interesting and applicable features (Woolley et al., 1968, Nedzinskas et al., 2012, Hazdra et al., 2008, Fali et al., 2014, Yekta Kiya et al., 2012, Azam Shafieenezhad, 2014). However, lasers may be used in very low (Le-Van et al., 2015, Rossetti et al., 2009, Tong et al., 2007) or high (Ohse, 1988, Rouillard et al., 2000) temperature conditions which will affect their work. Alongside, it is proved that temperature affects the lasing process through both change of the output photoluminescence and the laser characteristics (Rossetti et al., 2009) which are due to the dependence of carriers behavior on temperature.

In semiconductor (SC) hetero-structures containing two or more semiconductors with different lattice constants, band edge diagrams show more complexity than usual bulk semiconductors because of the important role of strain. Strain tensor depends on the elastic properties of neighbor materials, lattice mismatch, and geometry of the quantum dot (Trellakis et al., 2006).

This research will study the band structure and strain tensor of $In_xGa_{1-x}As$ quantum dots grown on GaAs substrate by quantum numerical methods to look for more efficient QDs by changing the working temperature of QD.

The rest of this paper is organized as follows: section II explains the model and method of the numerical simulation. Our results and discussions on the temperature effects are presented in section III. Finally, we make a conclusion in section IV.

## II. MODELING AND NUMERICAL SIMULATION

Band structure of a zinc-blende crystal can be obtained through:

$$\left(H_0 + H_k + H_{k.p} + H_{s.o.} + H'_{s.o.}\right)u_{nk}(r) = E_n(\mathbf{k})u_{nk}(r) \qquad (1)$$

In which $u_{nk}(r)$ is a periodic Bloch spinor (i.e., $\psi_{nk}(r) = e^{ik.r}u_{nk}(r)$) and

$$H_0 = \frac{p^2}{2m_0} + V_0\left(r, \varepsilon_{ij}\right) \qquad (2)$$

$$H_k = \frac{\hbar^2 k^2}{2m_0} \qquad (3)$$

$$H_{k.p} = \frac{\hbar}{m_0} \boldsymbol{k}.\boldsymbol{p} \tag{4}$$

$$H_{s.o.} = \frac{\hbar}{4m_0^2 c^2} \left( \boldsymbol{\sigma} \times \nabla V_0(\boldsymbol{r}, \varepsilon_{ij}) \right).\boldsymbol{p} \tag{5}$$

$$H'_{s.o.} = \frac{\hbar}{4m_0^2 c^2} \left( \boldsymbol{\sigma} \times \nabla V_0(\boldsymbol{r}, \varepsilon_{ij}) \right).\hbar\boldsymbol{k} \tag{6}$$

Here, $V_0(\boldsymbol{r}, \varepsilon_{ij})$ is the periodic potential of the strained crystal, $\boldsymbol{\sigma} = (\sigma_x, \sigma_y, \sigma_z)$ is the Pauli spin matrix, $c$ is the light velocity, and $m_0$ is the mass of electron. This equation can be solved by expansion of $V_0(\boldsymbol{r}, \varepsilon_{ij})$ to first order in strain tensor $\varepsilon_{ij}$ (Bahder, 1990). The resulting equation for a strained structure is then

$$\left( H_0 + \frac{\hbar^2 k^2}{2m_0} + \frac{\hbar}{m_0} \boldsymbol{k}.\boldsymbol{p}' \right) u_{nk}(\boldsymbol{r}) = E_n(\mathbf{k}) u_{nk}(\boldsymbol{r}) \tag{7}$$

where

$$\boldsymbol{p}' = \boldsymbol{p} + \frac{\hbar}{4m_0 c^2} \left( \boldsymbol{\sigma} \times \nabla V_0(\boldsymbol{r}, \varepsilon_{ij}) \right). \tag{8}$$

This equation can be solved by the second order non-degenerate perturbation scheme in which the latter terms are considered as the perturbation. Therefore, in the Cartesian space the solution is:

$$E_n(\mathbf{k}) = E_n(\mathbf{0}) + \frac{\hbar^2 k_i k_j}{2} \left( \frac{1}{m_n^*} \right)_{i,j} \tag{9}$$

In which the tensor of the effective mas is defined as:

$$\left( \frac{1}{m_n^*} \right)_{i,j} = \left( \frac{1}{m_0} \right) \delta_{i,j} + \frac{2}{m_0^2} \sum_{m \neq n} \frac{\langle n,0|p'_i|m,0\rangle\langle m,0|p'_j|n,0\rangle}{E_n(0) - E_m(0)} \tag{10}$$

and $i,j \equiv x,y,z$ (Galeriu and B. S., 2005).

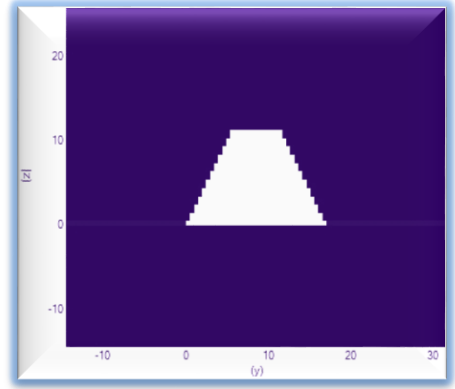

**Fig. 1:** *Cross-section of a pyramidal InGaAs QD with square base of the area $17nm \times 17nm$ and the height of 2/3 times the base width on $15nm$ thick GaAs substrate and $0.5nm$ wetting layer.*

In self-assembled $In_x Ga_{1-x} As$ QDs, firstly WL with a few molecular layers is grown on the substrate, and then, millions of QD islands grow on the wetting layer, each of which having a random shape and size. The resulting system is finally covered by GaAs. Many shapes can be approximated for QDs namely cylindrical, cubic, lens shape, pyramidal (Qiu and Zhang, 2011), etc. QDs are supposed to be far enough not to be influenced by other QDs. The one-band effective mass approach is used in solving the Schrödinger equation.

Fig. 1 shows the profile of a pyramidal $In_{0.4}Ga_{0.6}As$ QD surrounded by a substrate and cap layer of GaAs (Pryor, 1998). This ratio of indium is used in laser devices (Kamath et al., 1997). The structure of both GaAs and InAs is zinc-blende. The pyramid has a square base of area $17nm \times 17nm$ and the height of 2/3 times the base width on $15nm$ thick GaAs substrate and $0.5nm$ wetting layer. This structure is grown on (001) substrate index. As it can be observed from the picture, WL is much thinner than QD height. The growth direction of the structure is z.

The parameters related to the bulk materials used in this paper are given in Table 1, in which it is benefitted from references (Jang et al., 2003, Singh, 1993, Yu, 2010).

Also, for $In_xGa_{1-x}As$ the parameters are calculated as follows:

Lattice constant at T=300K (Adachi, 1983):

$$a = (6.0583 - 0.405(1-x)) \text{ Å} \qquad (11)$$

Effective electron mass at 300K (T.P.Pearsall, 1982):

$$m_e = (0.023 + 0.037(1-x) + 0.003(1-x)^2)m_o \qquad (12)$$

Effective hole mass at 300K (N.M., 1999):

$$m_h = (0.41 + 0.1(1-x))m_o \qquad (13)$$

Effective light-hole masses at 300K:

$$m_{lp} = (0.026 + 0.056(1-x))m_o \qquad (14)$$

Effective split-off band hole-masses at 300K is $\sim 0.15\ m_o$

### III. RESULTS AND DISCUSSION

Strain is generally defined as the value of length increase relative to initial length ($\varepsilon_L = \Delta L/L$) which in a better definition it can be the summation of all infinitesimal length increases relative to the instantaneous lengths ($\varepsilon_L = \sum \Delta L_t/L_t$). Therefore, taking into account length change in all directions, one achieves the tensor as:

$$\varepsilon_{ij} = \frac{1}{2}\left(\frac{du_i}{dr_j} + \frac{du_j}{dr_i}\right),\ \ i,j \equiv x,y,z \qquad (15)$$

where $du_i$ is the length change in i-th direction, and $r_j$ is the length in direction $j$ (Povolotskyi et al., 2004). The diagonal elements are related to expansions along an axis (stretch), but the off-diagonal elements represent rotations. This tensor is symmetric (i.e., $\varepsilon_{ij} = \varepsilon_{ji}$) although the distortion matrix $\boldsymbol{u}$ may be non-semmetric. In our case, the strain tensor is a diagonal matrix as follows:

$$\varepsilon = \begin{bmatrix} \varepsilon_{xx} & 0 & 0 \\ 0 & \varepsilon_{yy} & 0 \\ 0 & 0 & \varepsilon_{zz} \end{bmatrix} \quad (16)$$

in which $\varepsilon_{xx} = \varepsilon_{yy}$ due to the symmetry. This tensor shows a biaxial in-plain strain defined as:

$$\varepsilon_{xx} = \varepsilon_{yy} = \varepsilon_{||} = \frac{a_{||} - a_{substrate}}{a_{substrate}} \quad (17)$$

and a perpendicular uniaxial strain defined as (Peressi et al., 1998):

$$\varepsilon_{zz} = \varepsilon_{\perp} = -\frac{2C_{xy}}{C_{xx}}\varepsilon_{xx} = \frac{a_{\perp} - a_{substrate}}{a_{substrate}} \quad (18)$$

where $C_{ij}$ are components of the matrix which interconnects stress $\sigma$ to strain (i.e., $\sigma = C\varepsilon$) (Chuang and Chang, 1997). In Figs. 2(a) and 2(b) variation of strain is illustrated in different points of the middle cross-section of the structure. As it is seen, strain tensor is subjected to change in the both directions. However, obviously, the most variation is viewed at interfaces.

| Parameters used | GaAs | InAs |
|---|---|---|
| Band gap (0K) | **1.424eV** | **0.417eV** |
| lattice constant | **0.565325 nm** | **0.60583 nm** |
| Expansion coefficient of lattice constant | **0.0000388** | **0.0000274** |
| Effective electron mass (Γ) | **0.067m_o** | **0.026m_o** |
| Effective heavy hole mass | **0.5m_o** | **0.41m_o** |
| Effective light hole mass | **0.068m_o** | **0.026m_o** |
| Effective split-off mass | **0.172m_o** | **0.014** |
| Nearest neighbor distance (300K) | **0.2448 nm** | **0.262 nm** |
| Elastic constants | **C11 = 122.1** **C12 = 56.6** **C44 = 60** | **C11 = 83.29** **C12 = 45.26** **C44 = 39.59** |

***Table 1.*** *Parameters used in the model.*

In Fig. 3 the non-zero elements of strain tensor are drawn by going up through z-direction and at the middle of the structure. It can be noticed that the existence of indium in one side of interfaces has caused a jump in the strain tensor for $\varepsilon_{xx}$ ($\varepsilon_{yy}$) and $\varepsilon_{zz}$, meaning a tension in GaAs and compression in InGaAs lattice constants. This figure was found to be the same for different values of working temperature. This shows that change of temperature has no effect on the strain tensor.

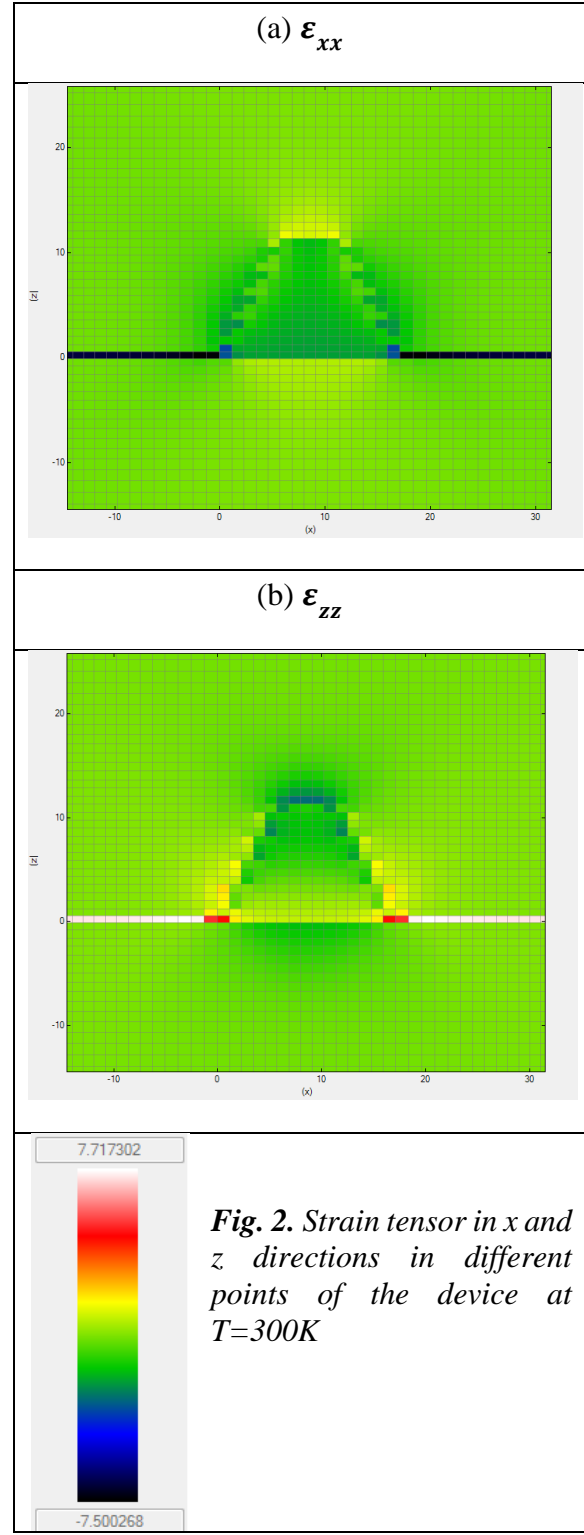

**Fig. 2.** *Strain tensor in x and z directions in different points of the device at T=300K*

This result seems logical, since as it was proved, strain is only dependent on the ratio of lattice constants which linearly change by temperature (i.e., $a = b(T - 300) + a_{300K}$). So their ratio remains fixed and strain does not change by temperature. This means that our result can be acceptable from analytical point of view.

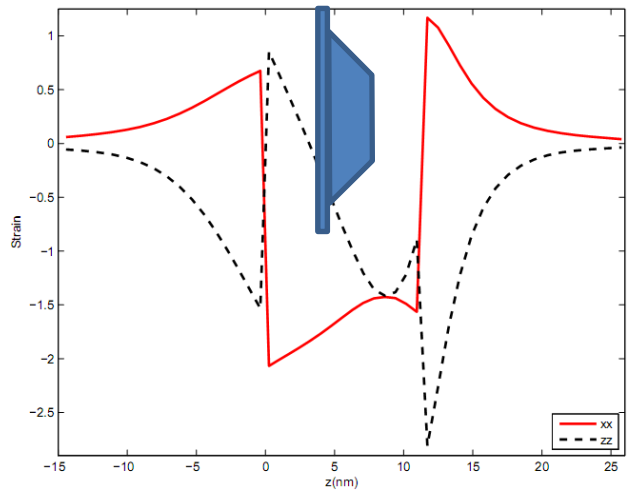

**Fig. 3.** *Nonzero elements of the strain tensor at $T = 300K$.*

Fig. 4(a,b,c,d) show the band-edge of each point of the of the pyramidal QDs at 300K for Γ, Heavy-Hole (HH), Light-Hole (LH), and Split-Off (SO) respectively. Both GaAs and InAs have direct band-gaps. As it is viewed, the band-edges change in different points. Fig. 4(a) indicates that deepest points into the QD have the most energy difference with the cap layer. Also, the cap layer band-edges have been also subjected to change in the points close to the QD. Also, Figs 4(b,c and d) show the hole band-edges which have a less values; nevertheless, their band-edges are different which can be explained as a result of changes in the effective masses, since heavy hole band has the largest, and split-off band has the smallest effective mass. All the figures show symmetry through x-axis. Moreover, the figures represent that band energy of the corners of the QD have matched their energy more than other points.

In Fig. 5 snapshots for conduction and valence Γ band-edges are shown in z-direction, and three first allowed energy states for electrons and holes are pointed out. All Heavy-Hole (HH), Light-Hole (LH), and Split-Off (SO) band-edges can also be seen in the valence band.

It is seen that in the mentioned physical conditions the first electronic eigenvalue separates from higher continuous states and lays down into the QD. This energy state is the ground state (GS) atom-like electron state with its special energy level which can be used in the recombination energy of electrons and holes. Other states are among continuous states of the GaAs.

However, the interesting phenomenon is the dependence of the banedges and energies to temperature. Obviously, by rise of temperature from 300K to 700K, the GS levels of electrons have displaced. The electronic states appear to decrease from 0.75eV to 0.55eV, showing that the e-h energy difference has been under change by temperature. This leads to different wavelengths of the QD laser. In the same way, it can be observed that all the band-edges and the band-gap have been temperature sensitive.

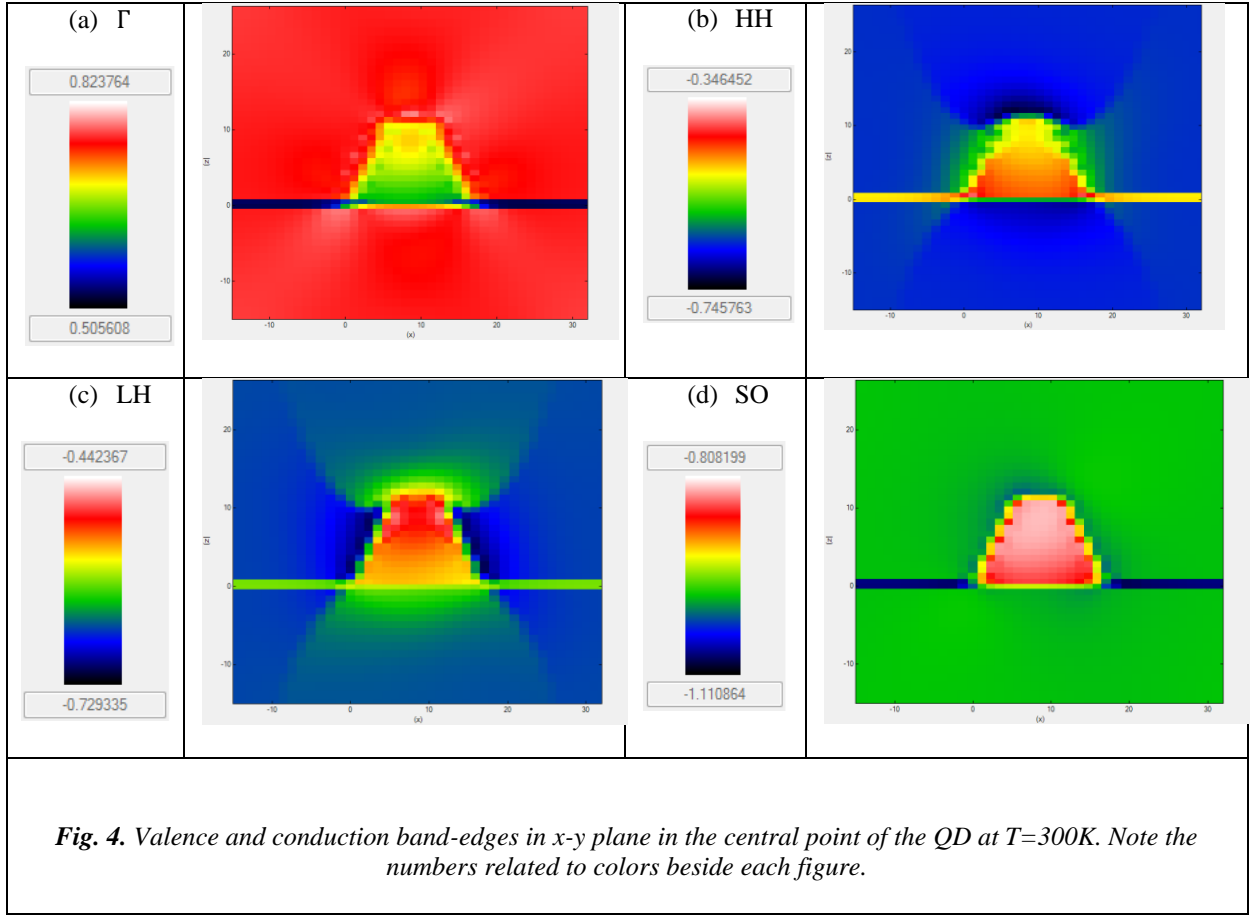

**Fig. 4.** *Valence and conduction band-edges in x-y plane in the central point of the QD at T=300K. Note the numbers related to colors beside each figure.*

Moreover, Fig. 6 illustrates a comparison of band edge for different values of temperature. Obviously, effect of temperature on the band-edges is remarkable in a wide range of temperature. As it is seen, rise of temperature has lowered the band-edges of both substrate and quantum dot.

Functionality of energy gap by change of temperature is shown to be

$$E_g = E_{g(T=0)} - 5.405 \times 10^{-4} \frac{T^2}{T + 204}$$

$$\& \ T \in [0,1000] \quad (19)$$

for GaAs, and

$$E_g = E_{g(T=0)} - 2.76 \times 10^{-4} \frac{T^2}{T + 83}$$

$$\& \ T \in [0,300] \quad (20)$$

for InAs, in which $E_g$ is in eV and T is in Kelvin (Fang et al., 1990). Also, for different ratios of indium in $Ga_x In_{1-x} As$, the behavior of energy gap is (Goetz et al., 1983):

$$E_g(x) = 0.36 + 0.63x + 0.43x^2$$

$$\& \ T = 300K \quad (21)$$

$$E_g(x) = 0.4105 + 0.6337x + 0.475x^2$$

$$\& \ T = 2K \quad (22)$$

The melting point for GaAs and InAs is respectively 1511K and 1215K. In Fig. 7 energy gap and recombination energy of the first eigenvalue are plotted as a function of temperature for a wide temperature range. This range is previously used for bulk materials in reference (Fang et al., 1990). Although all the temperatures are not meaningful, since the laser structure corrupts in hot conditions, the behavior can represent the whole effect of temperature. Interestingly, impact of temperature on energies is not linear, which is the important conclusion of this paper. Decrease of recombination energy results in the elongated laser wavelength which must be paid attention in the lasers.

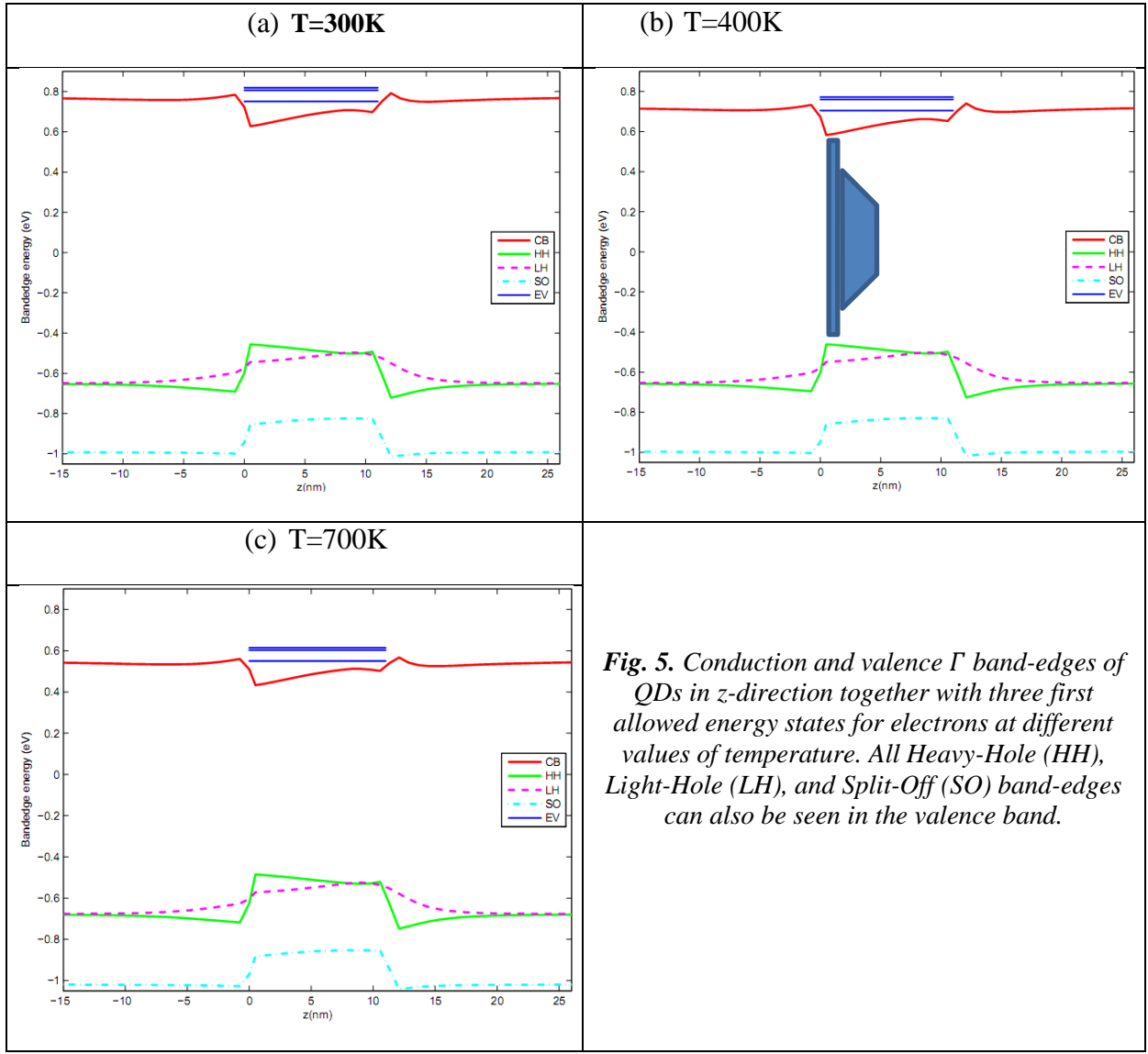

**Fig. 5.** *Conduction and valence Γ band-edges of QDs in z-direction together with three first allowed energy states for electrons at different values of temperature. All Heavy-Hole (HH), Light-Hole (LH), and Split-Off (SO) band-edges can also be seen in the valence band.*

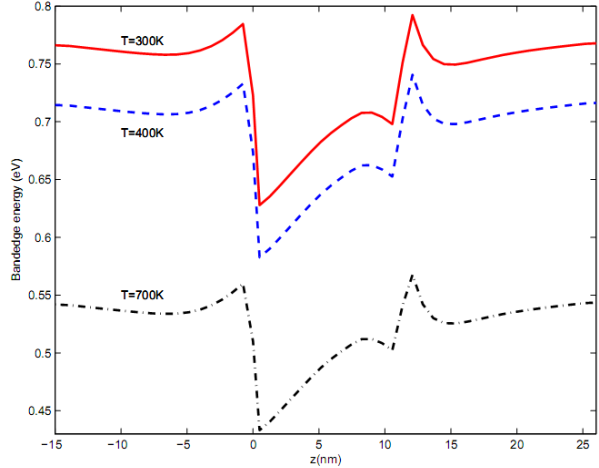

***Fig. 6.*** *Comparison of Γ band edges at different values of temperature.*

Comparison of these results which are related to bulk semiconductors show that our results can be logical. As it can be inferred, the behavior of our results is almost similar to the formula found in bulk samples.

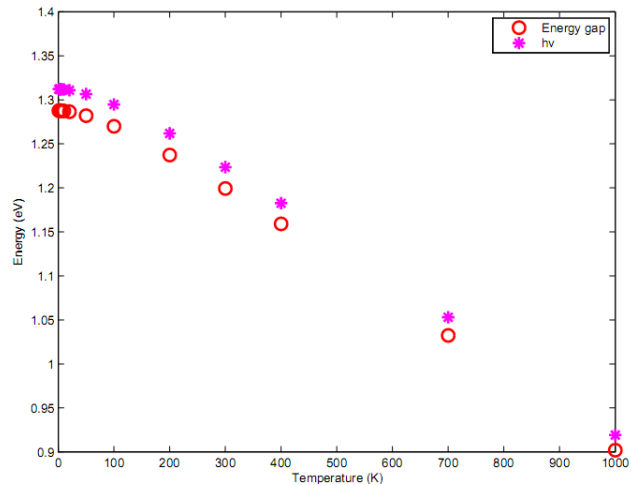

***Fig. 7:*** *Energy gap and recombination energy of the first electron- and hole-eigenvalues at different temperatures*

## IV. CONCLUSION

We investigated the band structure and strain tensor of $In_xGa_{1-x}As$ QDs grown on GaAs substrate by quantum numerical solution. It was shown that in usual temperatures, rise of temperature has no remarkable effect on strain tensor, but slightly decreases the conduction and valence band-edges as well as electron- and hole-energy states. Moreover, the main result was that temperature does not have a linear impact on the bandgap and recombination energies; both decrease by temperature rise.

## ACKNOWLEDGEMENT

The authors give the sincere appreciation to Dr. S. Birner for providing the advanced 3D Nextnano++ simulation program (Birner et al., 2007) and his instructive guides. We would like to thank numerous colleagues, namely Prof. S. Farjami Shayesteh, Dr. S. Salari, K. Kayhani, and Y. Yekta Kia for sharing their points of view on the manuscript.